\documentclass[11pt,preprint,aps]{revtex4-1}
\usepackage{helvet,times}
\usepackage{bm,textcomp}

\usepackage{setspace}
\usepackage{amsmath}
\usepackage{hyperref} 
\usepackage{graphicx} 
\usepackage[squaren,Gray]{SIunits}

\newcommand{\stension}{\lambda_s}			
\newcommand{\etension}{\lambda_e}			
\newcommand{\Rst}{R_{s}}					
\newcommand{\lf}{l_f}						
\newcommand{\Aadhesive}{A_{\mathrm{ad}}}		
\newcommand{\Aref}{A_{\mathrm{ref}}}
\newcommand{\dup}{\mathrm{d}}				
\newcommand{\nNpermum}{~\mathrm{nN / \text{\textmu} m}}	
\newcommand{\nNmum}{~\mathrm{nN \text{\textmu} m}}	
\newcommand{\nN}{~\mathrm{nN}}
\newcommand{\mum}{~\text{\textmu}\mathrm{ m}}
\newcommand{\VVianay}{ V_{el}}

\begin{document}

\setcounter{page}{1} 

\title{Dynamics of Cell Shape and Forces on Micropatterned Substrates\\
Predicted by a Cellular Potts Model}

\author{Philipp J.\ Albert and Ulrich S.\ Schwarz}

 \affiliation{Institute for Theoretical Physics and BioQuant, Heidelberg University, Heidelberg, Germany}

\begin{abstract}%
{Micropatterned substrates are often used to standardize cell experiments and to quantitatively study the relation
between cell shape and function. Moreover, they are increasingly used in combination with traction force microscopy on soft
elastic substrates. To predict the dynamics and steady states of cell shape and forces without any a priori knowledge of how
the cell will spread on a given micropattern, here we extend earlier formulations of the two-dimensional cellular Potts model.
The third dimension is treated as an area reservoir for spreading. To account for local contour reinforcement by peripheral
bundles, we augment the cellular Potts model by elements of the tension-elasticity model. We first parameterize our model
and show that it accounts for momentum conservation. We then demonstrate that it is in good agreement with experimental
data for shape, spreading dynamics, and traction force patterns of cells on micropatterned substrates. We finally predict shapes
and forces for micropatterns that have not yet been experimentally studied.}
\end{abstract}

\maketitle

\section{INTRODUCTION}

During attachment to a substrate, most cell types actively
sense the adhesive geometry and stiffness of their environment
by generating contractile forces in their actin cytoskeleton
that are transmitted to the substrate through cell-matrix
contacts \cite{Geiger2009}.
The resulting traction force then feeds back
into biochemical circuits of the cell by a large range of
different mechanosensitive processes, with dramatic consequences
for cell polarization, migration, proliferation,
differentiation, and fate \cite{Schwarz2012}. To understand these essential
processes, it is therefore very important to measure or predict
the cellular forces.

One of the biggest challenges in cell experiments is the
inherent variability in their organization, including shape
and traction forces. Cells on a homogeneously adhesive
substrate display a large variety of shapes, and even cells
with similar shapes usually differ in their internal organization.
To overcome this difficulty, micropatterned substrates
have emerged as a very useful tool to standardize
cell experiments \cite{Thery2006,vignaud_directed_2012}.
In a pioneering study using microcontact
printing, it has been shown that cell survival depends
also on the spatial extension of the pattern and not only on
the amount of ligand it contains \cite{Chen1997}.  
Many subsequent studies then used micropatterns to show that many essential cellular functions depend on shape, including the distribution
of stress fibers \cite{Thery2006a}, the orientation
of the mitotic spindle \cite{Thery2007} or endomembrane organization
\cite{Schauer2010}.

Cellular sensing of micropattern geometry is closely
related to stiffness sensing, as both depend on cellular forces
being developed in the actin cytoskeleton. To measure
cellular forces on flat elastic substrates, different variants
of traction force microscopy have been developed
\cite{dembo_stresses_1999,butler_traction_2002,Sabass2008}. This approach is now increasingly combined with micropatterning
of cell shape, for example, by using microcontact
printing \cite{Rape2011} or deep-ultraviolet illumination of polyacrylamide substrates \cite{Tseng2011} or lift-off techniques on silicone rubber substrates
\cite{hampe_defined_2014}.

Micropatterning of cell shape is naturally complemented
by quantitative image processing and modeling. Several
types of mathematical model have been developed to predict
cell shape on micropatterns  \cite{schwarz_physics_2013}.
The simplest type is a contour model. The simplest type is a
contour model. It has been suggested, based on observations
of circular arc features of cells adhering to homogeneous
substrates, that Laplace-type models arise from the competition
of tension in the periphery (geometrically a line
tension) and tension in the cell body (geometrically a
surface tension) \cite{Zand1989,Bar-Ziv1999}. 
Here, we call this approach the
simple tension model (STM). A quantitative analysis of
cell shape on dot patterns has shown that in the presence
of strong contour reinforcement by peripheral actin bundles,
the STM has to be modified by elastic elements, leading to
the tension-elasticity model (TEM) \cite{Bischofs2008}.
Both STM and TEM describe not only cell shape but also cell forces
 \cite{Bischofs2009}. It was shown recently that the TEM emerges as a
good approximation to a bulk model for contractile cells if
the tension in the periphery dominates the bulk tension
 \cite{farsad_xfem-based_2012,GuthardtTorres2012}. 
 
The natural starting point for a bulk model of cell shape
is continuum mechanics, which can be implemented with
the finite-element method (FEM). To represent contractility
in such a framework, one can use isotropic thermoelasticity,
which represents contractility by a negative pressure in
the elastic equations, as it can be induced in passive materials
by lowering temperature. This approach is commonly
used for model contraction in cell monolayers \cite{nelson_emergent_2005,Edwards2011,Mertz2012}.
Recently also, such an isotropic thermoelastic model
was used to predict the shape and forces of cells on micropatterns \cite{banerjee_controlling_2013}. To represent the anisotropic effect of
stress fibers, the isotropic thermoelastic approach has been
extended by an orientation-dependent order-parameter field
for contractility \cite{deshpande_bio-chemo-mechanical_2006,deshpande_model_2007}.
The strength of a stress fiber is
determined by a positive-feedback mechanism regarding
how much force can be built up in a given direction, favoring
directions of large effective stiffness. Cell shapes and
forces then can be predicted if the attachment sites are
known, for example, for micropatterns \cite{Pathak2008,farsad_xfem-based_2012}
and pillar arrays \cite{mcgarry_simulation_2009}.

FEM-based models for cell shape usually assume linear
elastic or hyperelastic material laws for the mechanical
properties of cells. Indeed, this is often a good assumption,
for example, on the large scale of tissues or for single cells
with conserved volume. However, for cells adhering to a
substrate, the projected area is not a conserved quantity,
and volume can be exchanged with the third dimension. In
this case, the mechanical response is mainly determined
by the actin cytoskeleton, which behaves like an elastic solid
under extension but does not resist compression because
actin filaments under compression can buckle, depolymerize,
and slide. When modeling cells on an intermediate
length scale, this fundamental asymmetry between tension
and compression can be represented by cable networks \cite{coughlin_prestressed_2003,Paul2008}. If actomyosin contractility is represented not by a
reduced resting length, but rather by a constant pull between
neighboring nodes, one arrives at the model of actively
contracting cable networks \cite{Bischofs2008,GuthardtTorres2012}.
Because contractile
tension dominates in the bulk and passive elastic tension
in the periphery, the corresponding computer simulations
are described well by the analytical predictions of the
TEM, both for actively contracting cable networks \cite{GuthardtTorres2012}
and thermoelastic continuum models \cite{farsad_xfem-based_2012}.

All of the above models are relatively static in nature
and assume that the general features of cell shape (in particular,
pinning points to the substrate) are already known.
Here, we aim to develop a model that predicts the dynamics
of cell shape and traction forces on micropatterned substrates
without any a priori assumptions regarding the final
shape. Two types of model in particular seem to qualify
for this purpose. Phase-field models (also known as levelset
methods) have recently been used to predict cell shape
in the context of cell migration, because they are particularly
suited to represent propagating contours
\cite{Shao2010,Ziebert2012,Ziebert2013}. 
To represent contractility, however, cellular Potts models
(CPMs) seem to be more appropriate. CPMs are latticebased
and represent a cell by a collection of spins (compare Fig.\ \ref{fig:Cell_cartoon} {\it a}),
thus allowing for arbitrary cell shapes. By defining
an energy functional on the spin configuration and exploiting
the slow timescale for cell spreading, one can use
Metropolis dynamics to propagate the system. CPMs are
commonly used to model tissues, as reviewed in Anderson
and Rejniak \cite{AndersonRejniak200708}. 
One prominent application is the study
of cell sorting by the differential adhesion hypothesis \cite{Graner1992,Glazier1993}. Besides their applicability to tissue,
CPMs have also been used in systems consisting of only a few
cells \cite{Kafer2007}. Recently, however, a CPM was applied also to predict the shape of single cells on micropatterned substrates \cite{Vianay2010}. Moreover, single-keratocyte movement has been modeled
with a CPM by coupling Metropolis dynamics to a model
for actin polymerization \cite{Maree2012}.

As in the case of the Ising model, the shapes predicted by
the CPM are dominated by interfacial tension. his generates a close relation not only to contour models, but also to vertex models. However, the latter are not lattice-based but rather define energy functionals for cell shapes with straight or circular contours (that is, for the solutions of the Laplace law). They have been used, for example, to
explore the role of mechanical interactions for growth of
the Drosophila wing imaginal disk \cite{Hufnagel2007} or to investigate
the influence of cell elasticity, cell-cell interaction, and
cell proliferation on cell sheet-packing geometry \cite{Farhadifar2007,aliee_physical_2012}.  However, vertex models cannot be used to model single cells on micropatterns because they cannot account for arbitrary shapes.

To arrive at a flexible and dynamic modeling framework
for cell shape and forces, here we choose to work with a
CPM. To predict not only shape, but also traction forces
of adhering cells, the CPM has to be modified in several
regards. For this purpose, we use insights from the TEM
to derive an energy functional for single cells on micropatterned
substrates. The energy functional is based on the
different kinds of tension acting in the cell, which are
balanced by the adhesive substrate and manifest as traction
force. From our predicted cell shapes, we can extract the
traction force for any adhesive geometry in a very efficient
manner and in good agreement with experimental results.

\section{CELL SHAPE MODEL}  

 \subsection*{Energy Functional}

\begin{figure}[t]
  \includegraphics[width=0.5\textwidth]{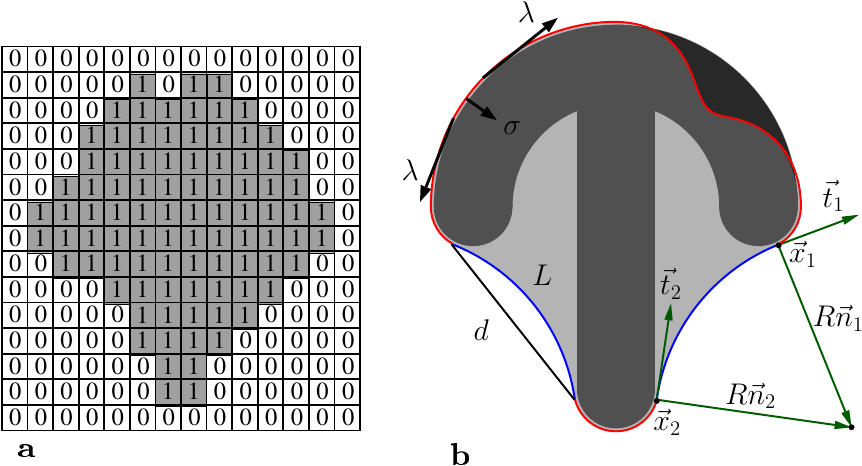} 
	\caption{\label{fig:Cell_cartoon} ({\bfseries\itshape a}) Representation of a cell in the cellular Potts model
(CPM). Occupied lattice sites are indicated by the index 1, empty sites
by 0. With a sufficiently large number of spins, the resulting shapes are
smooth and can be compared to other models or experiments. ({\bfseries\itshape b}) Schematic representation 
	of a cell ({\itshape light gray}) not fully spread on a crossbow-shaped micropattern ({\itshape dark gray}). Both the line tension $\lambda$ and the surface tension $\sigma$ act on the contour. 
	Free spanning arcs form at concave parts of the pattern. They are characterized by the spanning distance $d$ and arc contour length $L$. In the tension-elasticity model (TEM),
	the line tension $\lambda$ of the arcs is augmented by an elastic part.
	For the arc on the right side, the anchoring points of the contour are indicated by $\vec{x}_1$ and $\vec{x}_2$ with tangents $\vec{t_1}$ and $\vec{t_2}$.
	The normals $\vec{n}_1$ and $\vec{n}_2$ point to the center of the circular arc with radius $R$.}
  \end{figure}

Single cells on flat micropatterned substrates are effectively
two-dimensional and often form invaginated circular arcs
along free edges \cite{Bischofs2008}. The circular shape can be understood in context of a Laplace law, where a surface tension $\sigma$, which draws the contour inwards, is balanced by a line tension $\lambda$,
which wants to draw the contour straight
\cite{Zand1989,Bar-Ziv1999}. The surface tension results from the combined action of the plasma membrane wrapped around the cell
body and the actin cortex underlying it \cite{Lecuit2007}. In particular,
the actin cortex is contracted by myosin II minifilaments.
The line tension reflects the fact that the plasma membrane
and actin cortex are folded back onto themselves at the cell
periphery and thus lead to a geometrically different contribution
than in the bulk. Moreover, it reflects the fact that
actin filament bundles tend to accumulate in these folded
parts at the cell edge. The appearance of circular arcs is
not restricted to cells on a dot pattern; it also occurs on
concave parts of the micropatterned island \cite{Thery2006,Thery2006a}, as depicted for the crossbow shape in Fig.\ \ref{fig:Cell_cartoon}{\it
  b}. Here, the surface tension $\sigma$ pulls the contour inwards,
while the line tension $\lambda$ pulls the contour outwards.
This is different at the convex parts, where both surface tension
and line tension pull inwards. This pull is balanced by the
adhesion sites along the cell contour. Because the cell periphery
is less reinforced by actin bundles along the convex parts,
the line tension there is expected to be weaker.
    
The simplest model for the situation depicted in
Fig.\ \ref{fig:Cell_cartoon}{\it b} is the simple tension model
(STM), which assume both line tension $\stension$ and surface tension
$\sigma$ to be constant.  Then a Laplace law results that predicts
circular arcs with a constant radius $\Rst=\stension / \sigma$ to
occur along free parts of the contour. In experiments a correlation of
the arc radius $R$ and the spanning distance $d$ between two adhesive
islands was observed which can be understood in terms of the tension
elasticity model (TEM) \cite{Bischofs2008}. In this model, the
reinforced actin edge fibers forming at the free spanning edges
contribute with an elastic line tension $\etension=EA (L-L_0)/L_0$,
where the one-dimensional elastic modulus $EA$ accounts for the
rigidity of the contour and $L$ and $L_0$ are the contour length and
rest length, respectively (see Fig.\ \ref{fig:Cell_cartoon} {\it
  b}). Pinned parts of the contour are not reinforced and we assume
only the simple tension $\stension$ acting there. For the free arcs
one then has the overall line tension
$\lambda=\stension+\etension$. Like the STM, the TEM also leads to a
Laplace law. Combining this with the elastic line tension and the
geometrical relation between arc radius $R$ and spanning distance $d$,
one finds a self-consistent equation for $R(d)$:
  \begin{equation}
   \label{eq:RadiusSpanning}
   R=\lf \left( \frac{2R}{L_0} \arcsin\left(\frac{d}{2R} \right) -1 \right) + \frac{\stension}{\sigma},
  \end{equation}
where $\lf=EA/\sigma$ is a length scale defined by the relative weight of the one-dimensional elastic modulus and the surface tension. Therefore circular arcs also arise in the TEM, but with a radius $R$ that increases
with spanning distance $d$ and with contour reinforcement $\lf$ \cite{Bischofs2008}.
  
Like these contour models, the cellular Potts model (CPM) also centers
around the concept of geometrical tension, but allows for much more
variable geometries.  Our CPM implementation uses a two-dimensional
square lattice where the cell is represented by occupied lattice
sites, compare Fig.\ \ref{fig:Cell_cartoon} {\it a}.  The adhesive
pattern is realized by marking the corresponding lattice sites as
adhesive. Typically a pattern is made from 200 x 200 spins and the
cell is represented by 30.000 spins. The length $l$ of the total cell
interface is calculated with a modified marching square algorithm.
Because cell spreading is a relatively slow process (typical time
scale 10 min), one can assume that the system is close to mechanical
equilibrium and the Metropolis algorithm can be used to propagate cell
shape. During each step a lattice site at the periphery of the cell is
selected at random and inverted.  Then an appropriate energy
functional is used to calculate the energy difference $\Delta
H=H_{invert}-H_{current}$. The inversion is accepted with the
probability $e^{\Delta H/k_B T} $ if $\Delta H > 0$ and with certainty
otherwise.  Here the effective temperature $T$ governs the contour
fluctuation amplitude (typical value 0.2 in dimensionless units). Only
lattice sites at the periphery of the cells are chosen for update
attempts because cells do not form spontaneous holes in the
bulk or nucleate new material far away from the bulk. For a cell with
$n$ lattice sites in its periphery a Monte Carlo sweep is defined as
$n$ inversion attempts. A more detailed description of the
implementation can be found in the Appendix.

The core of the CPM is defined by the energy functional which we choose to be
  \begin{equation}
   \label{eq:Hamiltonian}
   H=\sigma A + \stension l + \sum_{\text{arc } i} \frac{EA}{2 L_{0,i}}(L_i-L_{0,i})^2 - \frac{E_0}{\Aref+\Aadhesive}\Aadhesive.
  \end{equation}
The first term accounts for the surface tension which scales with the
cell area $A$ as conjugated quantity. The second term is the
contribution of the simple line tension which scales with the cell
perimeter $l$. The third term is the sum over the contribution from
each actin edge fiber and a circle is fit to the corresponding part of
the contour to calculate $L_i$. 
All of the previously mentioned tensions contract a convexshaped
cell and are balanced by the adhesive geometry
accounted for by the fourth term. Cells form adhesive contacts
with the substrate, and the bond energy of each contact
lowers the total energy. The number of adhesion molecules
in a cell is finite and the energy gain by covering more
adhesive area therefore saturates with the covered area, $\Aadhesive$. This choice ensures a linear growth during initial
spreading, which later slows down and plateaus, as observed
for many different cell types
\cite{dubin-thaler_nanometer_2004,Cuvelier2007}. Strength and
saturation of the adhesive energy are controlled by $E_0$ and $\Aref$.
  
Cells in tissue are often described by CPM or vertex models with an
energy functional including an elastic (harmonic) constraint on the cell area
\cite{Graner1992,Glazier1993,Hufnagel2007,Farhadifar2007,Kafer2007,Vianay2010}. In
contrast to tissue, single cells on a substrate are essentially
two-dimensional and can increase their projected surface
area by taking material from the third dimension or by
making use of the excess area stored in the plasma
membrane or nearby vesicles. We therefore do not use an
elastic area constraint in our energy functional. The implications
of an elastic area constraint for the spreading dynamics
and for the dependence of arc radius on spanning distance
are discussed in the Appendix. It is shown there
that the third term in Eq.\ \ref{eq:Hamiltonian} can
also be interpreted as a saturation effect in membrane tension.
  
\subsection*{Parameter Estimation}
  
The surface tension has been estimated before as $\sigma\approx 2
\nNpermum$ from pillar deflections for endothelial cells
\cite{Bischofs2009} and as $\sigma\approx 0.7 \nNpermum$ from analysis
of the traction forces of epithelial cell sheets \cite{Mertz2012}.
The simple line tension should be
of the order of $\stension\approx 10\ \nN$, which is the typical force
acting on focal adhesion connected to the actin cytoskeleton
\cite{Balaban2001}. The rest length $L_0$ of the elastic arc is
assumed to be equal to the spanning distance, $d$.  The ratio
$\lf=EA/\sigma$ of elastic rigidity and surface tension describes the degree of arc reinforcement and, for computational simplicity, is taken to be constant, although in practice it might vary dynamically during cell spreading. It has been
estimated for buffalo rat liver cells cells on hard substrates with purely elastic arcs and a rest length of $L_0=1.01 d$ as $\lf=1300\ \mum$
\cite{Bischofs2008}. In our case, this value has to be reduced for several reasons, namely, that here we consider soft substrates, we have both simple and elastic tension, and we assume that  $L_0=d$. For typical arcs with
$R=15\mum$ and $d=12 \mum$, the same arc tension as in the purely
elastic case is reached for $\lf\approx340 \mum$ ($\sigma= 0.7
\nNpermum, \stension=10 \nN$). For a typical bundle radius of $100
nm$, this would correspond to a Young modulus in the MPa-range, as
found experimentally \cite{deguchi_tensile_2006}.

The two remaining parameters, $E_0$ and $\Aref$, can be estimated from
the adhesive energy density and the average cell size on homogeneous
substrates.  For weakly spread cells the adhesive energy gain in
Eq.\ \ref{eq:Hamiltonian} becomes $W=E_0/\Aref$ as the number of
adhesive contacts is not yet saturated. This adhesive energy density
reflects the amount of adhesion receptors available to the cell and
has been estimated before as $W=20 \nNpermum$ \cite{Cohen2004}.
Epithelial MCF10A cell on $3 \mathrm{kP}$ gels reach a spread area of
$A_0\approx 1700 \mum^2$ with an approximately round shape
\cite{Kostic2009}.  In order to relate these values to our model, we
note that there are no edge bundles on a homogeneous substrate and the
spread area, $A$, and adhesive area, $\Aadhesive$ in
Eq.\ \ref{eq:Hamiltonian} are equal. The energy functional
Eq.\ \ref{eq:Hamiltonian} depends then only on $A$ and its minimum
determines the final cell size.  From this we calculate
  \begin{eqnarray}
   E_0&=& W \Aref,  \\
   \Aref&=&A_0 \frac{\sigma \sqrt{A_0}+\stension \sqrt{\pi} +
						\sqrt{\sigma W A_0+ \stension W \sqrt{A_0 \pi}}}
					{(W-\sigma)\sqrt{A_0}-\stension \sqrt{\pi}}.
  \end{eqnarray}
From these formulae, we obtain typical values for the adhesion
parameters as $E_0=10^4 \nNmum$ and $\Aref=530 \mum^2$ $(\sigma=0.7
\nNpermum, \stension=10 \nN$).  In experiments, these values can be
varied for example by using micropatterns with different ligand
density or with mixtures of functional and denatured proteins.
    
  \subsection*{Traction Forces}
  
Contractile forces generated in cells are balanced by the adhesive substrate. Both the surface tension, $\sigma$, and the
simple line tension, $\stension$, pull normally to the contour, but the latter force depends on the curvature of the contour. For a given part of the contour with length $\dup l$ the force is \cite{Bischofs2009}
  \begin{equation}
   \dup \vec{F}= -(\sigma + \stension \kappa) \vec{n} \dup l,
   \label{eq:Traction_force}
  \end{equation}
  where $\kappa$ is the curvature and $\vec{n}$ the normal vector of unit length pointing outwards of the contour. Free spanning arcs are anchored at their endpoints and exert a force tangential to the contour \cite{Bischofs2009}
  \begin{equation}
	\vec{F}_{arc}=\etension \vec{t},
	\label{eq:Traction_force_arcs}
  \end{equation}
  where $\vec{t}$ is the tangent at the endpoints. Note that the edge bundles only exert a tangential force at their endpoints if their line tension $\stension+\etension$ is different from the value $\stension$ of the rest of the contour.
   
Having calculated the curvature and normal (as explained in the Appendix) of each lattice site for cell shapes predicted with our CPM, we can now use Eq.\ \ref{eq:Traction_force} to calculate the traction force acting on the adhesive part of the pattern beneath the contour. From circles fitted to the
free spanning arcs, the tangent at the anchoring points, arc
length, and spanning distance are estimated, which are
then used in Eq.\ \ref{eq:Traction_force_arcs} to calculate the force acting on the anchoring points (represented in the simulations by a single lattice site). Together with the force generated by the adhesive
part this gives the total traction force resulting from our
shape model.

For comparison with experimental results, several issues
have to be taken into account. First, our model predicts
spatially strongly localized forces, whereas in practice
they are typically distributed over a stripe of focal adhesions
along the cell contour 
\cite{Thery2006}. We therefore distribute our simulated forces on a
stripe of $2\mum$ width beneath the membrane using a disk shaped
kernel. Second, our model shows stochastic fluctuations in the cell
contour which are expected to be averaged out in experiments. We
therefore average it over $5 \times 10^5$ Monte Carlo sweeps to
account for the fluctuating contour due to the finite simulation
temperature. Third, traction forces are typically reconstructed in
experiments by an inverse procedure that filters out displacement
noise.  To obtain traction data which is comparable to experimental
results, we use the finite element method (FEM) as implemented in the
deal.II library to calculate the displacement from our CPM. In our FEM calculations, the forces generated by the cell are applied to the surface of a linear elastic material $250 \mum \times 250 \mum$ in area and $100 \mum$ deep with a displacement-free boundary at the bottom and stress-free boundaries at the sides. The hexahedral mesh is
locally refined beneath the cell. We then apply the Fourier-transform
traction cytometry method (FTTC)
\cite{butler_traction_2002,Sabass2008} to obtain a traction pattern
which can be compared directly with experimental results.  A complete
cycle of simulated forces, calculated displacements fields and
reconstructed forces can be found in the Appendix
(Fig.\ S4).
  
\subsection*{Momentum Conservation and Force Magnitude}

We now show that for our model the sum of all traction forces vanishes, as required by momentum conservation. 
For a contour $\vec{x}(l)=(x(l),y(l))$ parametrized by contour length $l$, the tangent is normalized and 
we can use Eq.\ \ref{eq:Traction_force} to write 
the sum of the traction forces resulting from the part of the contour
extending from $\vec{x}_1=\vec{x}(l_1)$ to $\vec{x}_2=\vec{x}(l_2)$ as
  \begin{eqnarray}
    -\int\limits_{l_1}^{l_2} (\sigma + \stension \kappa) \vec{n} \dup l 
    &=& -\int\limits_{l_1}^{l_2}  \left(\sigma \frac{\dup}{\dup l}   \binom{y}{-x} 
		- \ \stension \frac{ \dup \vec{t}}{\dup l} \right)  \dup l \nonumber \\
	&=& - \sigma M \Big(\vec{x}_2-\vec{x}_1\Big)+ \stension (\vec{t_2}-\vec{t_1}).
	\label{noarcresult}
  \end{eqnarray}
The minus in the second term comes from the normal pointing outwards. $M$ is the matrix for a $90^{\circ}$ counterclockwise rotation and $\vec{t_1}$ and $\vec{t_2}$ are
the tangents at the endpoints of the contour as illustrated in Fig.\ \ref{fig:Cell_cartoon} {\it b}. For cells without arcs, the start and end points of the integral are the same, and thus it vanishes, as required by momentum conservation.
  
For cells with a single arc, we can apply this calculation
only to the complementary part of the contour. Moreover,
we now have to account for the elastic line tension in the
arc according to Eq.\ \ref{eq:Traction_force_arcs}.
Because the arc is circular, the contour endpoints and circle normals are related by $\vec{x}_2-\vec{x}_1=-R (\vec{n}_2-\vec{n}_1)$.
Combining this with Eqs.\ \ref{noarcresult} and \ref{eq:Traction_force_arcs}, the total force  becomes
  \begin{equation}
  \int \limits_{l_1}^{l_2} \dup \vec{F} + \vec{F}_\mathrm{arc}
	  = \sigma R M (\vec{n}_2-\vec{n}_1) + (\stension +\etension) (\vec{t_2}-\vec{t_1}).
  \end{equation}
  Rewriting the arc radius in terms of tension, $R=(\etension+\stension)/\sigma$, and rotating the normals with the matrix, $M$, shows that the net force vanishes. 
  This also ensures momentum conservation in the STM as this simply corresponds to setting the elastic tension $\etension$ to zero. For cells with
  more than one arc, the same result follows by recursion.
  
  We finally comment on the magnitude of the total traction
force. For cells on a homogeneous substrate, this is simply
 \begin{equation}
	\int |(\sigma + \stension \kappa) \vec{n} | \mathrm{d}l	
	=\sigma l + 2 \pi \stension.
	\label{eq:total_force}
 \end{equation}
Thus, the total traction force scales linearly with cell perimeter $l$ as previously described for cell colonies \cite{Mertz2012}. Both shape and size of the cell change the perimeter $l$ and therefore influence the total force through the surface tension, $\sigma$. Larger cells or cells which deviate from a round shape exert a higher total force on the substrate than round cells with the same area, as it was found experimentally for rectangular micropatterns \cite{Rape2011}. 

\section{RESULTS}

\subsection*{Equilibrium Shapes and Cell Spreading}

\begin{figure}[t]
\centering
 \includegraphics[width=0.9\textwidth]{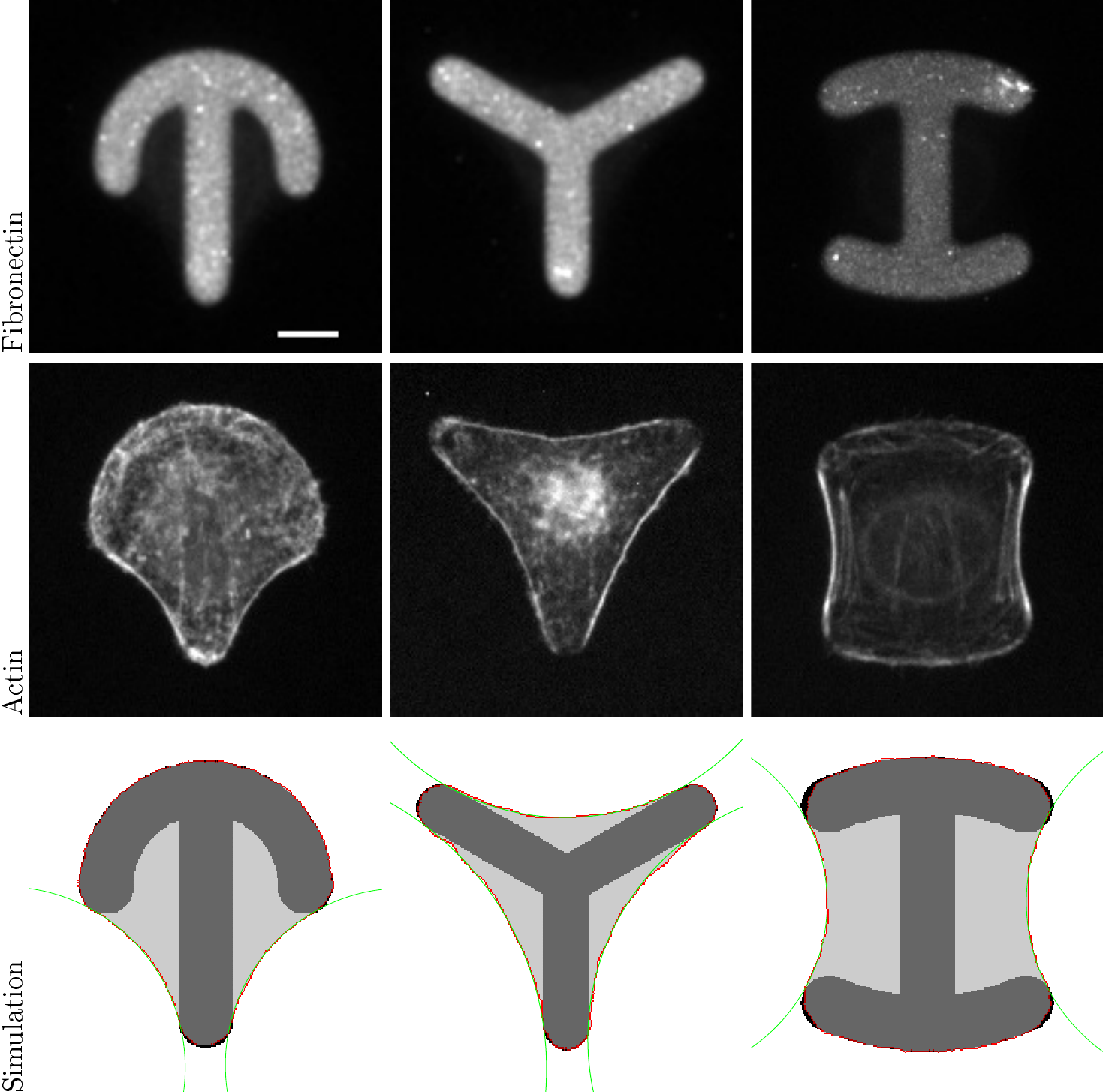}
 \caption{\label{fig:ShapePredictions} The top row shows experimental images for crossbow, Y and H patterns coated with fibronectin. Scale bar represents $10\mum$. The middle row shows HeLa-cells stained for actin on those patterns. The bottom row hows shape predictions by the CPM with circles fitted to the free spanning arcs. Experimental images kindly provided by Gintar\.{e} Garben\v{c}i\={u}t\.{e} and Vytaute Starkuviene-Erfle.}
\end{figure}

We first discuss the steady state shape of cells on micropatterns. In
Fig.\ \ref{fig:ShapePredictions} we compare experimental and simulated
shapes for HeLa cells plated on three fibronectin-coated
patterns commonly used for cell normalization, namely,
the crossbow, Y, and H patterns. The pattern width is equal
to $30\mum$ and the CPM
simulation uses a lattice constant of $0.15 \mum /\mathrm{pixel}$.
One can see that our model predicts very
well the typical sequence of convex and concave parts along
the cell contour. Moreover, the arc reinforcement modeled
by the TEM is clearly visible in all three actin images.
Note that here, the same parameter set ($\stension=10 \nN$,
$\sigma=0.7$, $l_f=340 \mum$, $A_0=1200\mum^2$, $E_0=7800 \nNmum$ and
$\Aref=390 \mum^2$) is used for all three cases,
because they have been realized on the same chip with the
same cell type and the same culture conditions.

\begin{figure}[t]
\includegraphics[width=1.0\textwidth]{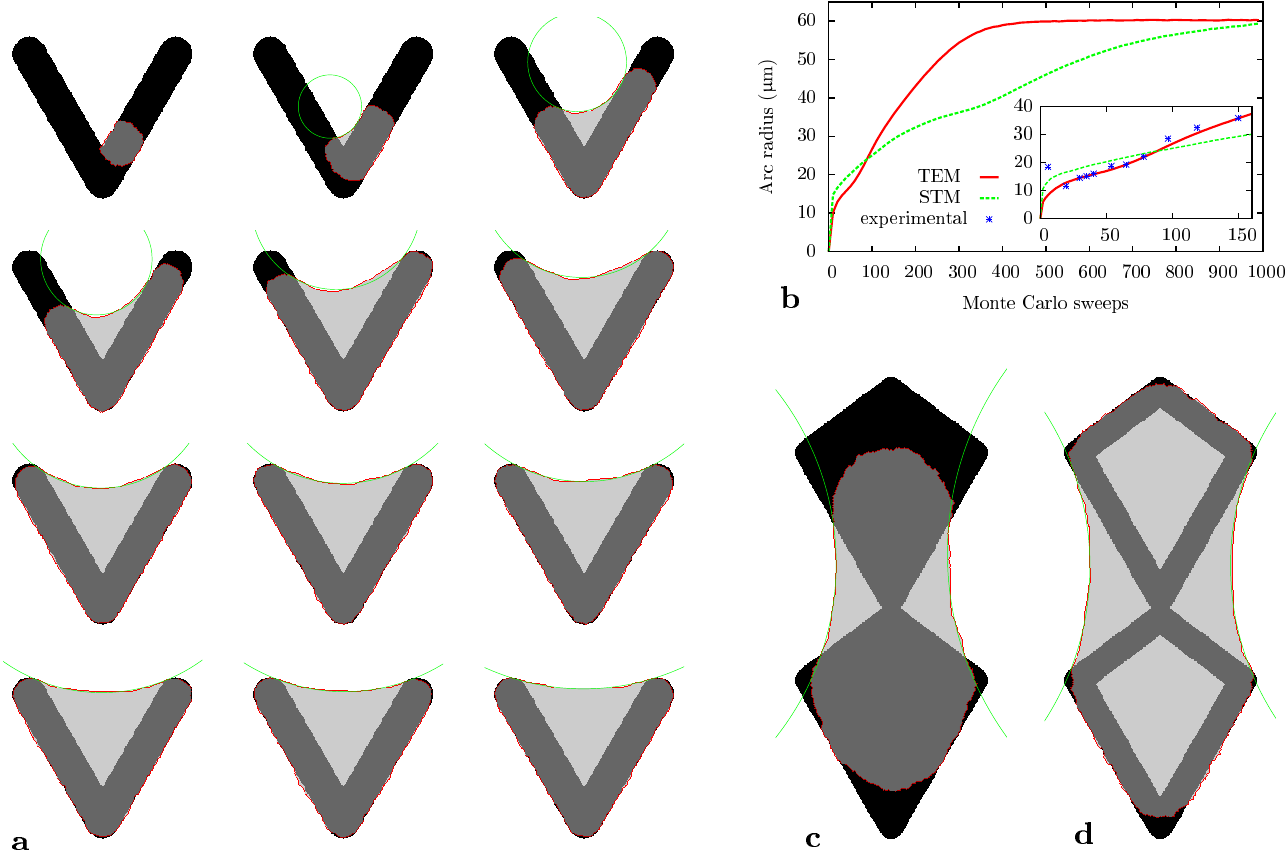}
 \caption{\label{fig:CellShapes} Cell shapes predicted by the CPM with surface tension $\sigma= 0.6 \nNpermum$, simple tension $\stension=10 \nN $, arc rigidity $EA=2000 \nN$, adhesive energy density $W=60 \nNpermum$ and cell target area of $A_0= 1700 \mum^2$. ({\bfseries\itshape a})  Cell spreading on ``V'' shaped pattern. The arms have an length of $46 \mum$ and the final spanning distance and radius are $d\approx 33 \mum$ and $R\approx 61 \mum$. A circle is fitted to the edge bundle. ({\bfseries\itshape b})  Radius of the circle fitted to the edge as function of Monte Carlo sweeps for cells described by the STM and TEM. The radius is averaged over $10^4$ cells all starting to spread at the same position as in the previous figure. Here, each Monte Carlo sweeps consists of $2 \times 10^4$ attempts to invert one of the boundary lattice sites. For the simulation of the STM cells a simple line tension of $\stension=36.6 \nN$ was used resulting in the same final radius as for the TEM cells. The inset shows the initial spread phase and data taken from \cite{Thery2006a}. ({\bfseries\itshape c})  Final cell shape on adhesive pattern which cannot be fully covered by the cell. Pattern has a with of $40 \mum$ and height of $96 \mum$. ({\bfseries\itshape d})  same as the previous figure but with a hollow adhesive geometry.}
\end{figure}

The steady state shapes shown in Fig.\ \ref{fig:ShapePredictions} result from a
dynamical spreading process that cannot be easily described
by standard models for cell shape. We now show that the
CPM also describes this process very well. In our simulations,
we start with a small spreading area of \texttildelow$10 \mum$ in
diameter at an arbitrary location on the pattern and then
let the cell spread. Here, we choose the V-shaped pattern
used in previous experimental studies \cite{Thery2006a}. For the RPE1
cells used in this study, the steady-state arcs have an unusual
small curvature, indicating that these cells have a weak bulk
contractility or very strong arc reinforcement. For both STM
and TEM, we first fix the model parameters such that this
final steady-state shape is achieved. We use a relatively
small surface tension of $\sigma= 0.6 \nNpermum$. In the STM, this surface tension requires a simple line tension of $\stension=36.6 \nN$ to reach the final arc radius of $61\mum$. Then, adhesive energies
of $W=60 \nNpermum$, several times larger than previous
estimates \cite{Cohen2004}, are required to allow spreading. 
For the TEM, the simple line tension can be reduced to $\stension=10 \nN$, because the final arc radius is also determined by the elastic contribution to the line tension. The required elastic tension of $EA=2000 \nN$ to match the final arc radius is similar to values reported previously for stiff substrates \cite{Bischofs2008}. In this case, an adhesive energy of $W=10 \nNpermum$ is sufficient to allow spreading.

With these parameters in place, we now can simulate the
spreading dynamics as shown in Fig.\ \ref{fig:CellShapes} {\it a} 
When the cells bridge the nonadhesive gap of the
V-shaped pattern, an actin edge bundle is formed, as indicated
by the green circle. After a main spreading phase,
during which the cell covers the complete adhesive area of
the pattern, the free spanning edge continues to move
outward, thereby increasing the spanning distance and the
radius of the edge bundle. The same two-step process is
seen in experiments \cite{Thery2006a}. The STM and TEM models differ strongly in the
timescale of spreading. Fig.\ \ref{fig:CellShapes} {\it b} shows the
radius of the circles fitted to the edge bundles averaged
over $10^4$
cells spreading on a V-shaped pattern. Cells
described by the TEM spread faster than cells described
by the STM. The reason lies in the increased simple line
tension of the STM needed to reach the same final radius
as the TEM. The inset of Fig.\ \ref{fig:CellShapes} {\it
  b} compares the curves for the initial spreading phase with experimental data \cite{Thery2006a}. TEM cells cover the whole adhesive area of the pattern after
80 Monte Carlo sweeps, whereas STM cells take up to 300
Monte Carlo sweeps. During this phase, the curves of the
two models are qualitatively similar. However, only the
TEM data can be fit well to the experimentally measured
data. This implies an important role of arc reinforcement
for the spreading process and sets the timescale to 30 min
for the 80 Monte Carlo sweeps. One can understand the
TEM as a mechanism that allows the cell to pull its contour
outward above nonadhesive parts of the substrate without
sacrificing any spreading potential above adhesive parts.
Spreading within the STM on the V-shaped pattern with
moderate choices for the simple tension and adhesive energy
would only be possible with a reduced surface tension.

In contrast to elastic continuum models, the CPM also
finds equilibrium shapes on adhesive islands that are too
large to be fully covered by the cell. This is demonstrated
in Fig.\ \ref{fig:CellShapes} {\it c}. Here, a tradeoff must be found between the
adhesive energy gain, which favors a large cell, the line
tension, which favors a small round cell, and circular arcs,
which should be as flat as possible to minimize the energy.
The result is a cell shape without sharp kinks and arcs
ending as parallel as possible to the pattern contour with a
large radius. In Fig.\ \ref{fig:CellShapes} {\it d} the same adhesive shape as in Fig.\ \ref{fig:CellShapes} {\it c} is used but with hollow diamond shapes. Now, the cell is able to cover the whole island because the reduction in adhesive surface leads to a less saturated adhesive energy gain in Eq.\ \ref{eq:Hamiltonian}. In addition, arcs form inside the diamonds while the cell is spreading giving the cell an overall concave shape with reduced area, as opposed to the partly convex cell on the filled diamond pattern in Fig.\ \ref{fig:CellShapes} {\it c}. 

\subsection*{Prediction of elastic substrate displacements}

\begin{figure}[htp]
\includegraphics[width=1.0\textwidth]{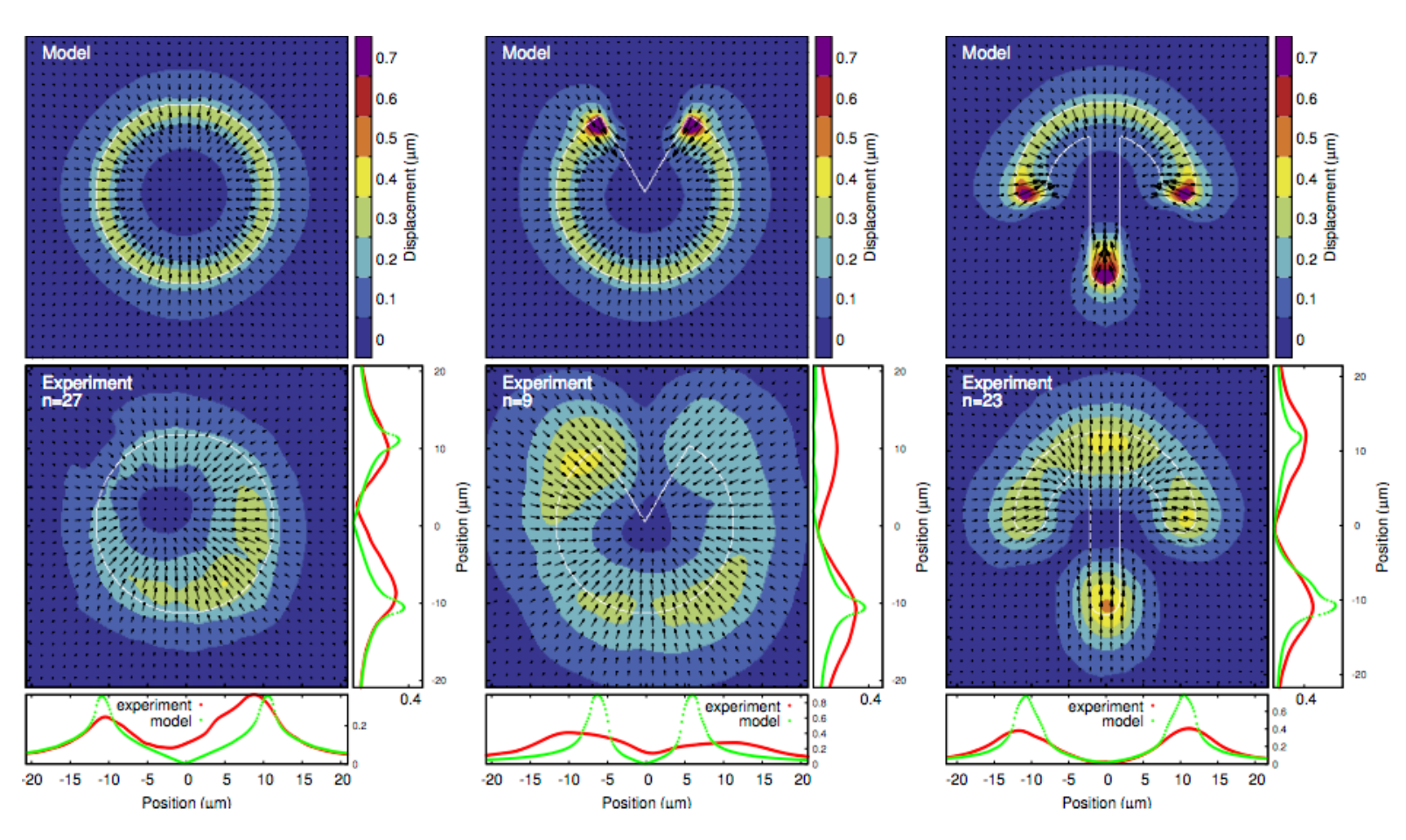}
\caption{\label{fig:BestFit05} Calculated displacements (upper panel) and experimental displacements (lower panel) for MCF10A-cells on fibronectin patterns on a polyacrylamide substrate with a Young modulus of $5 \mathrm{kPa}$
\cite{Tseng2011}. The smaller side panels are vertical and horizontal one dimensional slices of the displacement data. The slices always go through the center of the pattern except for the pacman shape where the horizontal slice goes through the tips. The number of averaged cells is indicated by $n$ for each pattern. Experimental data kindly provided by Martial Balland.}
\end{figure}
 
Although our final aim is to predict traction force patterns
for different micropatterns, we start our discussion of traction
patterns with displacement fields, because these are
the immediate outcome of experiments. Thus, we use the
FEM to calculate from our simulated forces the displacement
fields for an elastic substrate. We then compare these
results to experimental data and fit our three main model
parameters, namely, surface tension, $\sigma$, simple line tension, $\stension$, and elastic line tension $EA$. 
In Fig.\ \ref{fig:BestFit05} we show our results (upper panel) and compare them to experimental data (lower panel) \cite{Tseng2011}. The parameters  responsible
for the adhesive energy gain, $E_0$ and $\Aref$, are found to be unimportant for the fit quality.
Due to the small number of cells on the pacman pattern,
we exclude it from the fit.

Both the pacman and circular-disc patterns show the
localization of displacements at the cell contour, as predicted
by our model. For the circular disc, which only has
continuously adhesive edges, the displacements are directed
radially inward everywhere. This is different in the case of
the pacman pattern, where in both the simulation and the
experiment displacements deviate from the radial symmetry
at the tips of the wedge. The displacements there point
inward because of the actin edge bundle forming across
the wedge. Cells described by the STM would have displacements
pointing slightly away from the wedge because
of the curvature of the cell contour there. The fit of our
model parameters is dominated by the larger adhesive parts
of the contour for those patterns. The STM and TEM predict
similar magnitudes, which makes it difficult to distinguish
between the two. Displacements for the crossbow are dominated more strongly by the two actin edge bundles and
are largest at the extremities of the pattern due to the high
curvature and bundles originating there. The experimental
displacements in both arms of the crossbow are directed
more upward than expected from the model. This might
indicate the effect of internal actin fibers connecting from
the tips to the top of the micropattern. On the other hand,
displacements of the circular pattern and circular part of
the crossbow pattern are very similar in magnitude and
direction, indicating that polarized internal fibers only
play a minor role in this case.

To quantify the quality of the fit we use the norm
$L_2=\sum(u_{exp,i}-u_{sim,i})^2/\sum u_{exp,i}^2$, which is the
squared and normalized distance between the experimental and simulated
displacement fields. For the STM the fit yields a simple line tension
of $\stension = 5.53 \nN$, surface tension of $\sigma=0.56 \nNpermum$
and a $L_2$-value of 0.16. For the TEM the fit yields
$\stension=2.30\nN$, $ \sigma=0.83 \nNpermum$, $EA=40 \nN$ and a
$L_2$-value of 0.15. The TEM fits the data better, but
the difference is small, since both the disc and crossbow
are dominated by large convex parts where actin edge bundles
are unimportant.

The TEM decreases the simple line tension while
increasing the surface tension. As discussed above for
spreading on the V-shaped pattern, the TEM allows the cell to pull its contour outward above nonadhesive parts of
the substrate with a reduced simple tension. The increased
surface tension in the TEM compensates for the reduced
simple tension, but the contribution of the surface tension
is curvature-independent. Thus, the TEM allows smaller
forces in regions of high curvature while keeping forces in
small-curvature regions the same as in the STM.

The overall agreement of our model with the experimental
displacements demonstrated in Fig.\ \ref{fig:BestFit05} is quite good. However, the experimental displacements decay
more slowly than the simulated ones, for several possible
reasons. First, the $2 \mum$ wide stripe beneath the membrane
where we apply forces to the substrate might be too narrow,
and the size of the adhesions might vary with force
\cite{Balaban2001,Trichet2012}. The agreement with the experimental data can be improved by increasing the stripe width to $4 \mum$, but this does not
appear to be reasonable given the actual size of the adhesions
and the feature size of the micropattern. It is therefore
more likely that the disagreement arises from the limited
resolution of bead tracking and from the registration and
averaging procedures, both of which blur the displacement
fields. Variations in the pattern shapes from manufacturing
or by deformation from the cells also widen the force spots
and make pattern registration more difficult. Because these
experimental limitations might be improved in the future, for our theoretical predictions here we keep the
$2 \mum$ scale for the adhesion width.

\subsection*{Prediction of traction forces}

\begin{figure}[t]
\includegraphics[width=1.0\textwidth]{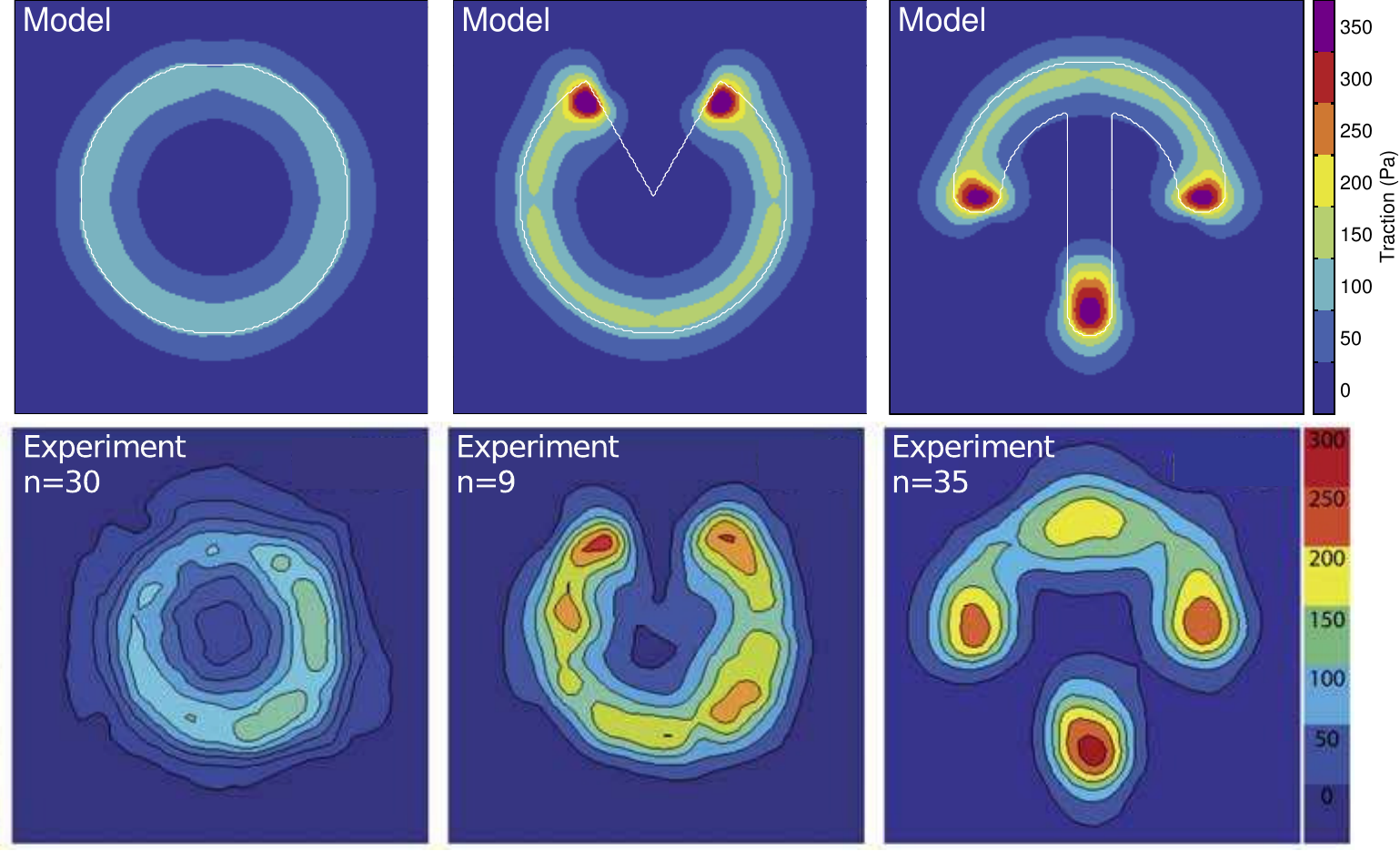}
 \caption{\label{fig:BallandReconstruction} The top row shows reconstructed traction forces for the TEM with the best fit parameters ($\stension=2.30\nN$, $ \sigma=0.83 \nNpermum$, $EA=40 \nN$) and the bottom row traction forces
reconstructed from experimental data for MCF10A-cells on fibronectin patterns on a polyacrylamide substrate. Experimental data reproduced from \cite{Tseng2011} with permission of The Royal Society of Chemistry.}
\end{figure}

Given the displacement data discussed above, one can now
reconstruct traction forces that resemble those obtained
from experimental data. For this purpose, we use FTTC
with regularization \cite{butler_traction_2002,Sabass2008}. In the top row of
Fig.\ \ref{fig:BallandReconstruction} we show the traction force reconstructed from the simulations shown in Fig.\ \ref{fig:BestFit05}.
A comparison with the experimental data (Fig.\ \ref{fig:BestFit05}, lower row) \cite{Tseng2011} shows that our procedure predicts most of the experimental features. The only exception
seems to be the additional localization of experimental
traction forces in the upper part of the crossbow pattern,
which might be due to the occasional presence of internal
stress fibers along the long side of this pattern.

\begin{figure}[t]
   \includegraphics[width=.95\textwidth]{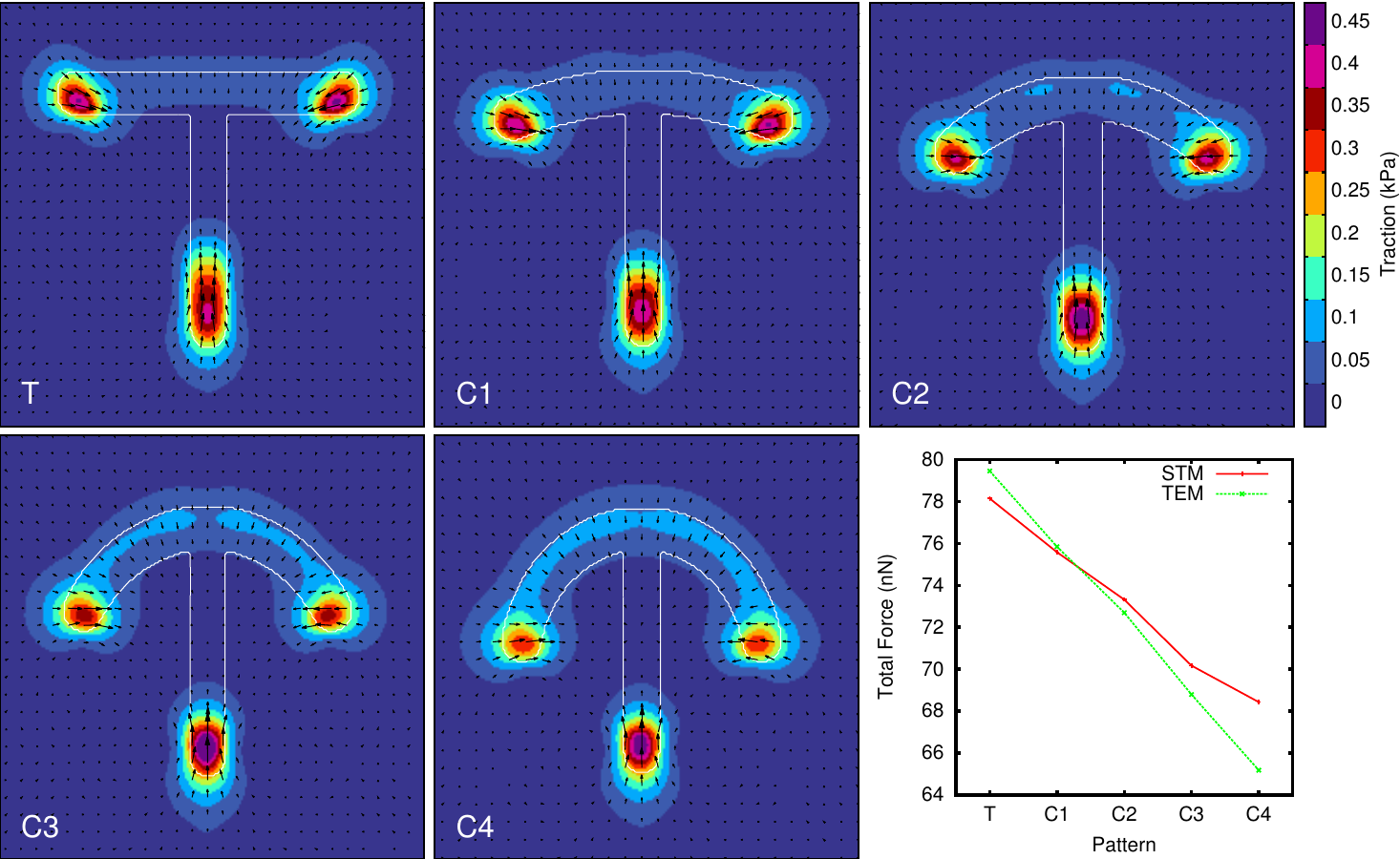}
 \caption{\label{fig:CurvatureChange} Reconstructed traction force for T-shaped pattern changing the curvature of the top bar into a crossbow shaped pattern. Cells are modelled with the TEM with $\sigma=0.65 \nNpermum,\lf=184 \mum,\stension=2.7 \nN$. The final panel shows the total force for the different curvatures both for TEM and STM cells with $\sigma=0.5 \nNpermum, \stension=6.5 \nN$. The pattern widths are $\mathrm{T}=32 \mum,\mathrm{C1}=30.8 \mum,\mathrm{C2}=29.34 \mum,\mathrm{C3}=27.49 \mum,\mathrm{C4}=25 \mum$ and ensure the same cell area on all patterns.
 }
\end{figure}

The influence of cell shape on force generation is best
seen by gradually changing the adhesive geometry. Both
curvature and spanning distance of the free arcs are varied
in Fig.\ \ref{fig:CurvatureChange} as the T-shaped pattern is transformed into a
crossbow for cells described by the TEM. For all shapes,
the forces are localized to the extremities of the patterns,
but increasing the curvature relocalizes them from the end
points to the adhesive edge of the contour, and at the same
time the force direction in the prominent force spots changes
from being aligned with the edge bundles to a more radial
orientation. Both observations, the force increase in the
adhesive contour and the orientation change, have been observed for
RPE1 cells \cite{Vignaud2013}. The increase of forces with
higher curvature at the adhesive contour is a consequence
of the simple line tension acting in the contour (compare
Eq.\ \ref{eq:Traction_force}). The prominent force spots in the extremities are
due to the larger curvature and the edge bundles being
anchored there. Changing the curvature of the top bar also
decreases the spanning distance, and therefore the force in
the bundles, from $7.75 \nN$ in the T-shaped pattern to $4.33 \nN$ in the final crossbow, which explains the decrease
of forces in the prominent spots and also the directional
change, since arc forces become less important. The force
spot at the bottom of the T-shaped pattern is less localized
compared to that in the final crossbow, because the edge
bundles do not always attach to the outermost part of this
pattern.

\begin{figure}[t]
 \includegraphics[width=1.0\textwidth]{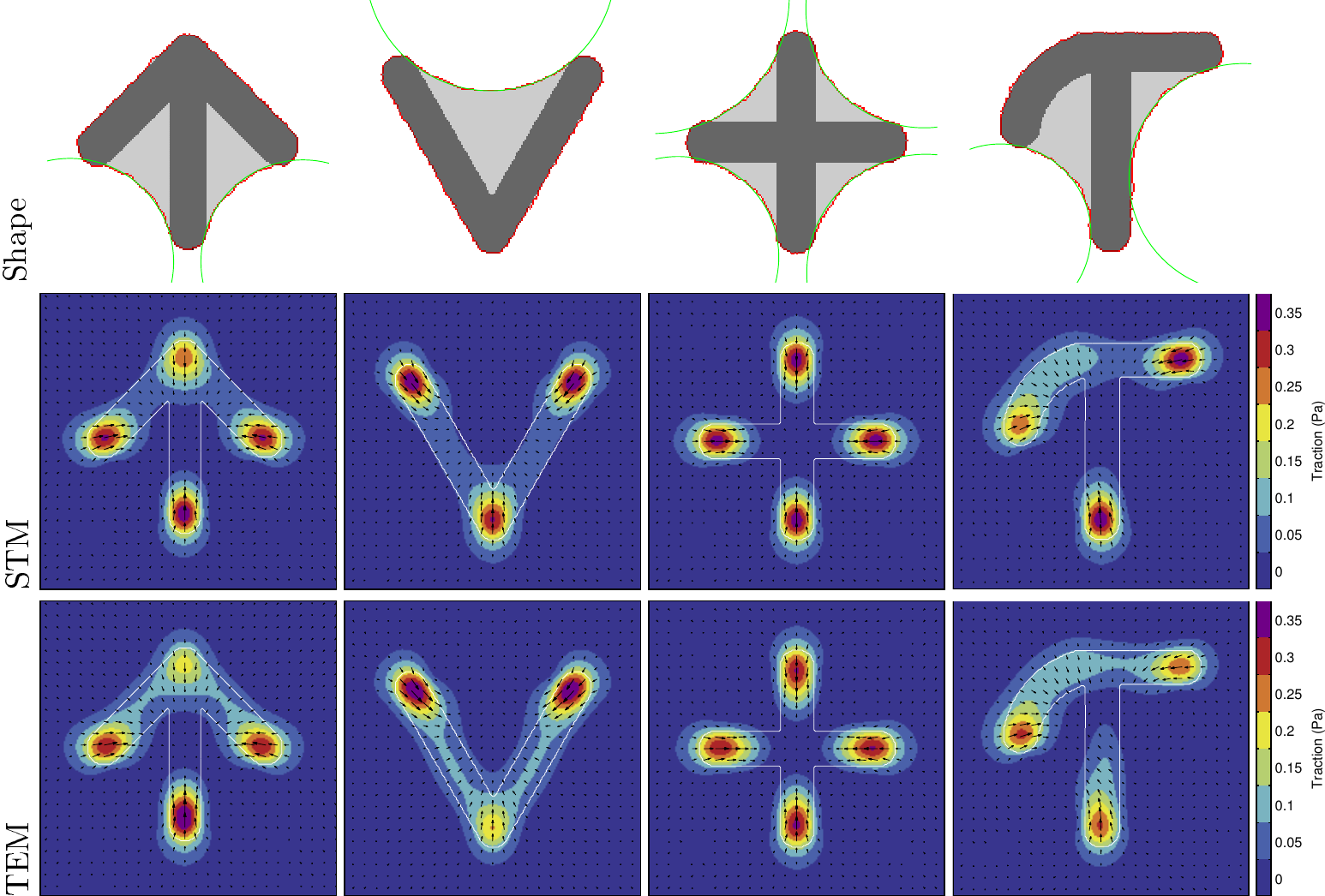}
  \caption{\label{fig:Supp_Shapes_Traction_Further_Geometries}
    Predicted shapes and traction forces for further adhesive
    geometries, both for STM and TEM. All patterns have a width of $25
    \mum$. The parameters are taken from the best fit to the
    displacement data from Fig.\ \ref{fig:BestFit05}. STM with $\sigma=0.56 \nNpermum,
    EA=0,\stension=5.53 \nN$. TEM with $\sigma=0.83 \nNpermum,EA=40
    \nN,\stension=2.3 \nN.$}
\end{figure}

The STM yields similar results for the total force, but
forces are always pointing radially inward. The decrease
of the total force in the STM is a consequence of the rounder
shape of the cells on the crossbow pattern as stated by Eq. \ref{eq:total_force} (the pattern dimensions are chosen to ensure the same cell area on all patterns, but their perimeter decreases linearly).
For cells described by the TEM, the total force is also
influenced by the decreased bundle tension, as reflected by
the steeper slope in the {\textit lower right} panel in
Fig.\ \ref{fig:CurvatureChange} which shows the total force as
function of curvature. In
Fig.\ \ref{fig:Supp_Shapes_Traction_Further_Geometries} we show a gallery of additional shape and traction predictions
for micropatterns for which no traction force fields
have yet been reported in the experimental literature.
Because these simulations are computationally very cheap, the CPM approach is a very helpful exploratory tool
for quickly assessing the effect of a newly designed
micropattern.

\section{DISCUSSION}

In summary, the CPM is a versatile tool for robust prediction
of cell shapes and forces on any micropattern of interest.
The underlying reason is that it actually models the dynamic
process of shape determination. Thus, it also makes predictions
on spreading dynamics and does not require any a
priori knowledge of the final shape. As shown above, all
of our predictions agree quite well with published experimental
data.
To adapt the CPM for the prediction of shapes and forces
of single cells on two-dimensional micropatterns, we have
made several modifications to the original formulation of
the CPM model for cell shape in tissue. Despite its large
contribution to the current understanding of cell sorting
and cell shapes in tissue \cite{AndersonRejniak200708}, the
CPM has been used before only a few times to describe single cells
\cite{Vianay2010,Maree2012}. To further
advance this approach, we have added two essential
elements to the conventional CPM formulation. Motivated
by the TEM, we have added an elastic line tension to
describe the effect of contour reinforcement of edge bundles.
Moreover, we have added an adhesion term to the
energy functional that does not constrain cell area but acts
like a reservoir for additional area that can be used if sufficient
ligand is present. The good agreement of our model
with experimental data, demonstrated here, confirms that
our energy functional describes the main features of this system.
One of the biggest advantages of our approach is that it
is computationally inexpensive (typical runtime is on the
subsecond timescale), thus making it an ideal exploratory
tool for quickly establishing typical cell behavior on micropatterns
without any a priori knowledge of the final shape.

Our model reveals that the TEM model allows for faster
cell spreading within a reasonable parameter range and
that elastic arcs act to relieve tension from adhesive parts
of the contour while maintaining the same cell shape. The
TEM makes spreading above nonadhesive parts easier and
at the same time allows the cell to generate traction forces
more by bulk than by contour tension. Our CPM assumes
that spreading is limited mainly by the availability of adhesion
receptors, thus leading to saturation in adhesive area.
This approach neglects other potential limitations of
spreading, most notably the effect of increased membrane
tension \cite{gauthier_temporary_2011}.As shown in the Appendix, to first order the energy functional of
Eq.\ \ref{eq:Hamiltonian} does not make a fundamental
distinction between these two limitations for
spreading. However, appropriately designed micropatterns
might in fact be an appealing way to investigate these
important questions in the future, both in experiments and
in the framework of the CPM.

Combining the CPM with a contour model allowed us to
interpret the energy terms in the CPM energy functional as tensions and provided an easy way to predict traction forces.
The contour model also connects our three model parameters,
namely, simple line tension, elastic line tension,
and surface tension, to the cell geometry via the relation
between spanning distance and arc radius
(Eq.\ \ref{eq:RadiusSpanning}). For
each cell type of interest, our model parameters can be fitted
to the experimentally observed cell shape on a reference
pattern and then used to predict cell shape and forces on
other patterns. As more information becomes available for
the detailed molecular structure of actin cortex and peripheral
fibers, our model can be modified to include such information,
e.g., by replacing the TEM term in the energy
functional of Eq.\ \ref{eq:Hamiltonian} with a more detailed expression.

There are two important aspects of cell adhesion to
micropatterns that are not addressed in this study. First,
our model does not describe the effect of internal structures
like nonperipheral stress fibers.We expect that this is a good
approximation, as long as several cells are averaged to
obtain a generic result that averages out individual inhomogeneities.
Otherwise, more detailed models are required that
also include internal stress fibers \cite{loosli_cytoskeleton_2010}.  Such extensions might profit from recent advances in our understanding of
the internal structure and dynamics of different kinds of
stress fibers \cite{hotulainen_stress_2006,tanner_dissecting_2010}. Second, our model does not explicitly
describe the effect of the mechanical feedback between
elastic substrate and cell adhesion, which allows the cell to sense the rigidity of its environment \cite{Geiger2009,Schwarz2012,guvendiren_stiffening_2012}. 
To include this important aspect in our model, it had to be extended by
models of the mechanosensitive organization of the adhesion
structure and the actin cytoskeleton
\cite{shemesh_physical_2012}. At the current
stage, the effect of stiffness is incorporated by fitting the
model parameters to experimental reference data and calculating
displacement fields with the correct rigidity values of
the substrate.

Here, we have focused on spreading dynamics and
steady-state properties of sessile cells. In this case, the
Metropolis approach is expected to work well, because the
cell is essentially relaxing to local mechanical equilibrium.
When combined with a model for actin polymerization
\cite{Maree2012}, our model can be extended in the future to study also persistent cell movement on single micropatterns or on networks
of micropatterns.

\begin{acknowledgments}
{We thank J\'{e}r\^{o}me Soin\'{e} and Christoph Brand for help with the FEM- and
optimization software and Sebastien Degot, Yoran Margaron and Michel Bornens for helpful
discussions regarding cells on micropatterns.
We thank Gintar\.{e} Garben\v{c}i\={u}t\.{e} and Vytaute Starkuviene-Erfle from Heidelberg
for providing the experimental images shown in Figure 2 and Martial Balland from Grenoble
for providing the experimental displacement data used in Figure 4. 
The authors acknowledge support by the EU-program MEHTRICS.
USS is a member of the Heidelberg cluster of excellence CellNetworks.}
\end{acknowledgments}



\begin{appendix}
\setcounter{figure}{0}

\renewcommand{\thefigure}{S\arabic{figure}}   

\bibliographystyle{biophysj}

\section{Radius Spanning Distance Relation with Elastic Area Constraint}

Many cellular Potts models (CPM) \cite{Graner1992,Glazier1993,Vianay2010} or
vertex models \cite{Hufnagel2007,Farhadifar2007,Kafer2007} for cells in tissue use
an elastic (harmonic) constraint for the cell area or volume in combination with
a simple line tension mediating cell-cell interaction. In addition,
some models also include an elastic line
tension \cite{Farhadifar2007,Kafer2007}. For a single cell the
simplest energy functional combines a simple line tension with the
elastic area constraint.
  \begin{equation}
   \label{eq:Hamiltonian_with_area_constrain}
   E=\stension l + k(A-A_0)^2,
  \end{equation}
where the first term accounts for the simple line tension $\stension$
which scales with the cell perimeter $l$ and the second term describes
the area elasticity with elastic coefficient $k$ and target area $A_0$.

\begin{figure}[t]
  \includegraphics[width=\textwidth]{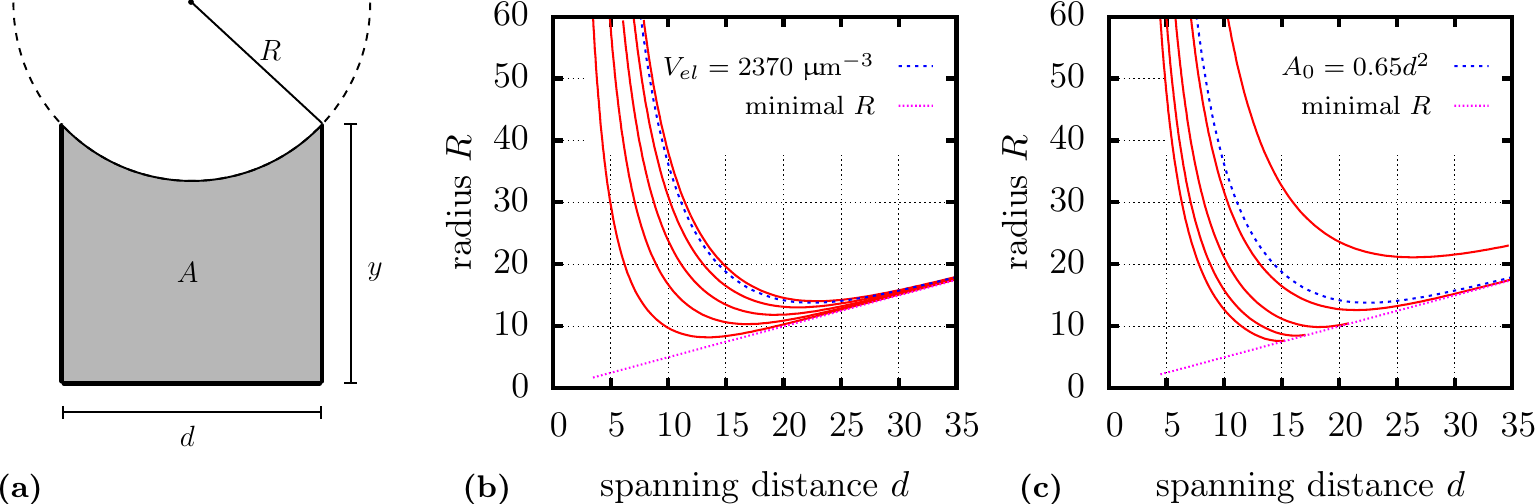}
  \caption{\label{fig:LagrangeArea}(a) Cell with area $A$ confined to
    a U-shaped micropattern with a single circular arc. (b) $R(d)$
    dependence calculated from Eq.\ \eqref{eq:R(d)_area_constrain}
    with $\VVianay=\stension/k$ increasing from $\unit{500}{\micro
      \metre^{-3}}$ (bottom curve) to $\unit{2500}{\micro
      \metre^{-3}}$ (top curve) and a target area of
    $A_0=0.65d^2$. (c) $R(d)$ dependence for a target area increasing
    from $A_0=0$ (bottom curve) to $A_0=0.8 d^2$ (top curve) and
    $\VVianay=\unit{2370}{\micro \metre^{-3}}$. The blue dashed curves
    are the same in both figures with $\VVianay=\unit {2370}{\micro
      \metre^{-3}}$ and $A_0=0.65 d^2$ both taken from
    \cite{Vianay2010}. }
 \end{figure}

As in the tension-elasticity model (TEM), the dependence of the arc
radius $R$ on the spanning distance $d$ can be found by a force
balance. First we define surface tension $\sigma=\partial E / \partial A$ and
line tension $\stension=\partial E / \partial l$. The surface tension
pulls inwards perpendicular to the contour and the line tension exerts
a force depending on the curvature of the contour. The force balance
then reads
  \begin{equation}
  \label{eq:Force_balance}
  \sigma \vec{n}=\stension \frac{\mathrm{d}{\vec{t}}}{\mathrm{d} l}= \stension \frac{1}{R} \vec{n},
 \end{equation}
where the contour is parameterized by its length $l$, and $\vec{n}$ and $\vec{t}$ are the normal and the tangent to the contour, respectively.
Calculating the derivative of Eq.\ \eqref{eq:Hamiltonian_with_area_constrain}, Eq.\ \eqref{eq:Force_balance} becomes
 \begin{equation}
  2 R (A-A_0) - \frac{\stension}{k}=0.
  \label{eq:Force_balance_with_area}
 \end{equation}
This shows that the cell area $A$ and the arc radius $R$ are not
independent of each other. In contrast to the TEM or simple-tension
model (STM), the arc radius now is not controlled locally, but
depends on the overall cell shape. Area changes at one end of the cell
can influence the arc radius at the other end. For a cell on a
U-shaped micropattern as depicted in Figure \ref{fig:LagrangeArea}a,
the implicit equation for the arc radius is found from
Eq.\ \eqref{eq:Force_balance_with_area} as
 \begin{equation}
	 2 R \left(\frac{d}{4} \sqrt{4 R^2-d^2}  - R^2 \arcsin\left(\frac{1}{2} \frac{d}{R}\right)+ dy - A_0 \right) - \frac{\stension}{k}=0.
	 \label{eq:R(d)_area_constrain}
 \end{equation}
 As in the TEM the arc radius depends on the spanning distance $d$,
 but in addition also the height $y$ of the U-shaped pattern
 influences the radius. In the following only the case of quadratic
 shapes with $y=d$ is considered.

There are two modes of controlling the cell shape, either by changing
the target area $A_0$ or by changing the ratio of simple tension and
the strength of the area constraint $\VVianay=\stension/k$.  Figure
\ref{fig:LagrangeArea}b shows $R(d)$ for different values of
$\VVianay=\stension/k$. The monotonously increasing relation between
arc radius and spanning distance observed in experiments
\cite{Bischofs2008} can be achieved by shifting $\VVianay$ to very
small values. However, this brings the radius very close to the
minimal possible radius of $R=d/2$ and such strongly invaginated cells
are usually not observed experimentally. In Figure
\ref{fig:LagrangeArea}c the target area $A_0$ is changed. For small
target areas Eq.\ \eqref{eq:R(d)_area_constrain} does not have a
solution for all spanning distances and the $R(d)$ curve ends at the
minimal radius condition. Cells would collapse in such
geometries. Figure \ref{fig:LagrangeArea}b and \ref{fig:LagrangeArea}c
also show $R(d)$ (blue curves) for parameters used previously
\cite{Vianay2010}. Neither changing the strength
$\VVianay=\stension/k$ nor changing the target area $A_0$ can achieve
a monotonously increasing relation between arc radius and spanning
distance without yielding strongly invaginated cells. The addition of
an elastic line tension as in the TEM to the energy functional
\eqref{eq:Hamiltonian_with_area_constrain} does not change this
outcome.

\section{Spreading Dynamics}
\begin{figure}[t]
\begin{center}
  \includegraphics[width=0.6\textwidth]{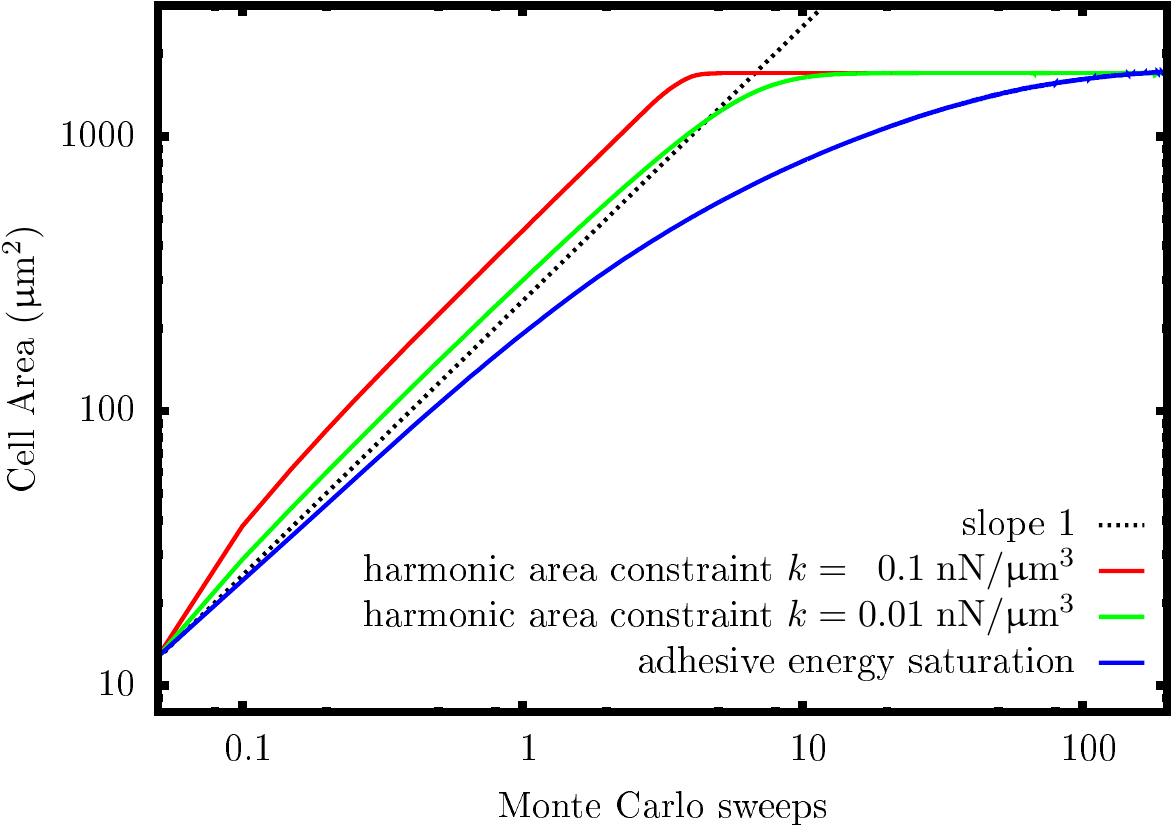}
\end{center}
  \caption{\label{fig:RadiusSpreading} Cell radius during spreading as
    function of Monte Carlo sweeps for an elastic area constraint (see
    Eq.\ \eqref{eq:Hamiltonian_with_area_constrain}) and an saturation
    of the number of adhesion molecules (see Eq.\ 2 in the main
    text). The simulations are carried out with a simple line tension
    of $\stension=\unit{10}{\nano \newton}$, elastic rigidity as
    stated in the legend, target area $A_0=\unit{1700}{\micro
      \metre^2}$. For the model with saturation of adhesive energy the
    same parameters as for Figure 3 of the main text where used
    ($\sigma=\unit{0.6}{ \nano \newton}$, $W=\unit{60}{\nano \newton
      \per \micro \metre}$, one Monte Carlo sweep consists of $2\times
    10^4$ inversion attempts).}
\end{figure}

The spreading dynamics of a cell with an elastic area constraint as in
Eq.\ \eqref{eq:Hamiltonian_with_area_constrain} and a cell with a
saturation of the number of adhesion molecules as in Eq.\ 2 of the
main text is expected to be different. For the former the quadratic
area constraint yields a large contribution throughout the spreading
process as long as cell area $A$ and target area $A_0$ do not
match. For the latter the energy gain from forming new adhesion
contacts stays within the same order throughout the whole spreading
process. This is reflected by the simulated cell area as function of
Monte Carlo sweeps shown in Figure \ref{fig:RadiusSpreading}. For an
elastic area constraint the linear area growth stops only shortly
before the target area is reached. Within the Metropolis dynamics all
steps increasing the cell size are accepted due to the large
contribution of the area term. The growth cannot be faster than
linear since only single lattice sites are inverted during each step. In a
model where the cell spreads against a viscous force the area would
initially grow stronger than linear. With a saturation of the adhesive
energy the growth gradually slows down similar to what is seen in
experiments \cite{Gauthier2011,Cuvelier2007}. Reducing the strength
$k$ of the elastic area constraint makes the transition to the steady
state less abrupt, but it also results in larger fluctuations around
the target area.

We also note that the limitation in adhesive area used here is similar
to a limitation in membrane tension. For a cell on a
homogeneously adhesive substrate the cell area and adhesive area are
equal $A=\Aadhesive$ and no actin edge bundles exist. With the
adhesive energy density $W=E_0/A_{ref}$ the
energy functional Eq.\ (2) of the main text becomes
\begin{align}
 E&=\lambda_s l+\sigma A - \frac{E_0}{A_{ref}+A}A \nonumber  \\
 &=\lambda_s l+\sigma A - \frac{E_0}{A_{ref}+A}A - W A +W A  \nonumber \\
 &=\lambda_s l + \underbrace{ \left[ \sigma + \frac{W A}{A_{ref}+A} \right] }_{ \sigma'}A - WA
\end{align}
where $\sigma'$ now takes the role of the surface tension. It
increases with the cell area which can be interpreted as an increase
due to a finite amount of membrane area. Since the
energy functional stays essentially the same, the spreading dynamics
and the force measurements are not directly affected by the
different interpretations of the energy functional.

\section{Implementation}
  
\begin{figure}[t]
  \includegraphics[width=\textwidth]{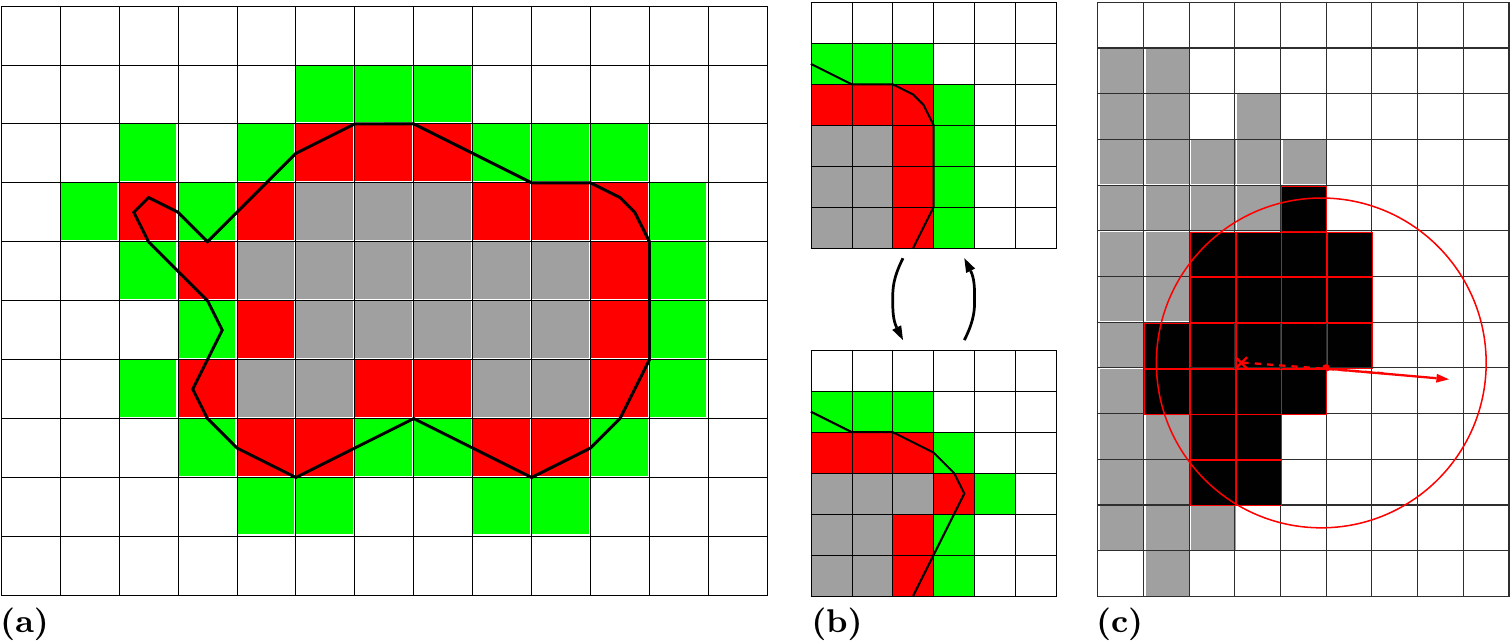}
\caption{\label{fig:Cell_on_lattice} Cell represented on a
  lattice. Sites inside the cell are gray, sites in the border layer
  of the cell are red and sites not being occupied by the cell but in
  the neighborhood of border sites are green. The cell periphery is
  indicated by the black line. (a) Representation of a whole cell. (b)
  Update of a single lattice site.(c) Circular mask applied to the
  cell boundary to define the normal.}
  \end{figure}

In the lattice-based CPMs a cell is represented by a set of occupied
lattice sites as illustrated in Figure
\ref{fig:Cell_on_lattice}a. Under normal conditions it is very
unlikely that cells form spontaneous holes or that new part of the
cell nucleate without contact to the bulk. We therefore use a modified
Metropolis algorithm \cite{Glazier1993} which only allows to invert
sites at the cell boundary.
  
The red sites in Figure \ref{fig:Cell_on_lattice}a are occupied by the
cell and form the outermost border layer of the cell. They are
surrounded by the outside layer of sites (green) not occupied by the
cell but being adjacent to one of the border sites. Updates by the
Metropolis algorithm only happen in these two layers. The gray lattice
sites belonging to the cell bulk are passive (no holes are formed), as
are the white medium sites (no nucleation away from the
bulk). Occupied lattice sites can get isolated from the bulk of the
cell when the cell retracts.

To define the cell boundary we use the marching square algorithm, a
two-dimensional variant of the marching cube algorithm
\cite{Lorensen1987}. Given four lattice sites the marching square
algorithm defines the orientation and length of the boundary between
those four sites. To calculate the contribution of a single lattice
site to the cell perimeter the occupation of its eight surrounding
sites needs to be known. The marching square algorithm allows only
boundary orientations of $0 ^{\circ},45 ^{\circ},90 ^{\circ},\dots$,
which results in a high lattice anisotropy. The anisotropy can be
reduced by allowing more possibilities for the boundary orientation
and length by taking a larger neighborhood. Each occupied lattice
sites gets a value assigned increasing with the number of occupied
neighbors. A higher value pushes the boundary further away from this
site. Similar methods to refine the marching square algorithm have
been used in image processing \cite{Mantz2008}.  We use a square of
4x4 lattice sites to define length and orientation of the boundary
between the four central sites. The boundary contribution of a single
site is then defined by its 24 neighbors. The resulting cell outline
is shown in Figure \ref{fig:Cell_on_lattice}a as black line. The
length of this line is used as the cell perimeter in the energy
functional Eq.\ (2) of the main text. The cell area is defined by the
number of occupied lattice sites.

Figure \ref{fig:Cell_on_lattice}b illustrates the inversion of a
lattice site. A new site is added to the green outside layer in this
case. The changes required by the inversion in the red border and
green outside layer are stored in a lookup table which requires
knowledge about the identity of the four surrounding sites. The
change in circumference calculated by the refined marching square
method is also stored in a lookup table requiring the occupation
values of the 24 surrounding sites.

The orientation of the normal to the cell border is found by applying
a circular mask to the lattice as illustrated in Figure
\ref{fig:Cell_on_lattice}c. The vector connecting the geometrical
center (cross) of the occupied lattice sites within the circle (black sites)
and the circle center defines the normal
direction $\vec{n}$ at the circle center. With this normal the
boundary segments left and right of the original segment and their
normal orientations $\phi_l$ and $\phi_r$ can be identified. The
curvature is then approximated by
  \begin{equation}
   \kappa=\frac{1}{2} \frac{\phi_l-\phi_r}{l},
  \end{equation}
where $l$ is the length of the boundary segment. The factor 1/2 arises from two boundary segments sharing one kink in the boundary.

\section{Reconstruction of Traction Force from Simulated Force Fields}
\begin{figure}[t]
 \includegraphics[width=0.9\textwidth]{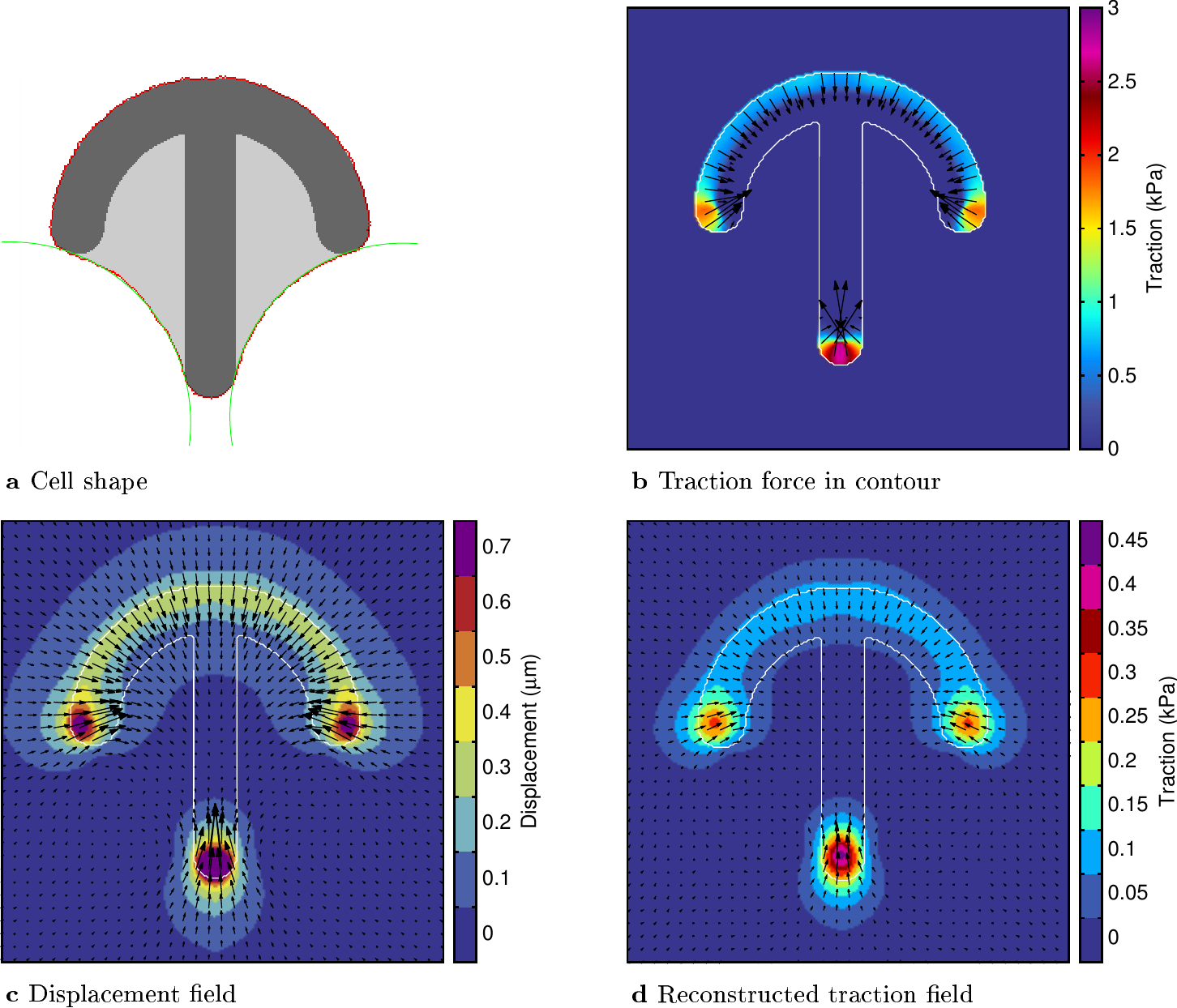}
 \caption{\label{fig:Supp_Workflow} Different stages of the reconstruction of traction forces from the CPM.}
\end{figure}

As described in the main text, the simulated traction force is used to
calculate a displacement field from which the traction can be
reconstructed with the Fourier-transform traction cytometry method
\cite{Sabass2008} to make it comparable to experimental
results. First, the cell shape is obtained as shown in Figure
\ref{fig:Supp_Workflow}a. Due to the finite simulation temperature the
membrane fluctuates and the traction force in the cell contour is
obtained by averaging over $2 \times 10^5$ Monte Carlo sweeps. The
traction force is distributed over a $\unit{2}{\micro \metre}$ wide
stripe beneath the membrane. To be more precise, we apply a disk
shaped kernel with a radius of $\unit{2}{\micro \metre}$ to each
lattice site distributing the forces to the surrounding sites. In
combinations with a pattern-shaped kernel it is ensured that traction
forces are only applied to adhesive parts of the patterns and that
forces from membrane fluctuation above non-adhesive parts are
propagated to the pattern. Both magnitude and vector field of the
traction force are shown in Figure \ref{fig:Supp_Workflow}b. Figure
\ref{fig:Supp_Workflow}c shows the displacement field found by our
finite element method for the traction field in Figure
\ref{fig:Supp_Workflow}b on a substrate with Young modulus of
$\unit{5}{\kilo \pascal}$. From the displacement field the traction
force is reconstructed with the Fourier-transform traction cytometry
method with a regularization parameter of $3 \times 10^{-8} \micro
\metre^2 \per \pascal^2$ arriving at the final result shown in Figure
\ref{fig:Supp_Workflow}d. Our choice of the regularization parameter
yields the same total traction before and after reconstruction.

\end{appendix}

\end{document}